\documentclass[12pt]{iopart}
\begin{document}
\input epsf
\newcommand{\ttau}{r}
\newcommand{\vev}[1]{\langle #1 \rangle}
\def\mapright#1{\!\!\!\smash{
\mathop{\longrightarrow}\limits^{#1}}\!\!\!}
\newcommand{\bigoint}{\displaystyle \oint}
\newlength{\extraspace}
\setlength{\extraspace}{2.3mm}
\newlength{\extraspaces}
\setlength{\extraspaces}{2.5mm}
\newcounter{dummy}
\newcommand{\be}{\begin{equation}
\addtolength{\abovedisplayskip}{\extraspaces}
\addtolength{\belowdisplayskip}{\extraspaces}
\addtolength{\abovedisplayshortskip}{\extraspace}
\addtolength{\belowdisplayshortskip}{\extraspace}}
\newcommand{\ee}{\end{equation}}
\newcommand{\figuur}[3]{
\begin{figure}[t]\begin{center}
\leavevmode\hbox{\epsfxsize=#2 \epsffile{#1.eps}}\\[3mm]
\bigskip
\parbox{15.5cm}{\small \ 
\it #3}
\end{center} \end{figure}\hspace{-1.5mm}}
\newcommand{\fig}{{\it fig.}\ }

\title{On RG-flow and the Cosmological Constant.}

\author{Erik Verlinde}

\address{ Physics Department, Princeton University, Princeton, NJ 08544}

\begin{abstract}
From the AdS/CFT correspondence we learn that the effective action of
a strongly coupled large $N$ gauge theory satisfies the Hamilton-Jacobi
equation of 5d gravity. Using an analogy with the relativistic point
particle, I construct a low energy effective action that includes the Einstein 
action and obeys a Callan-Symanzik-type RG-flow equation.
The flow equation implies that under quite general conditions
the Einstein equation admits a flat space-time solution,
but other solutions with non-zero cosmological constant are also allowed.
I discuss the geometric interpretation of this result in the context of
warped compactifications. This contribution is an expansion of
the talk presented at the conference and is  based on work 
reported in \cite{one} and \cite{two}.

\end{abstract}




\section{Introduction}

The problem of the cosmological constant involves high energy as well as low 
energy physics. It is not just sufficient to have a zero cosmological 
constant at energies near the Planck scale, one also needs to 
explain the absence of vacuum contributions
at much lower energy scales. This low energy aspect of the cosmological
constant problem is the most puzzling, and seems to require a fundamental 
new insight in the basic principles of effective field theory, the 
renormalization group and gravity.
String theory/M-theory, in whatever form it will eventually be formulated,
has to provide a solution to this problem if it wants to earn a status as
a true fundamental theory of our Universe. This
resolution may depend on specific details of the underlying 
fundamental theory, but, whatever the mechanism is at high energies, 
it must manifest itself in some way in low energy terms as well.

So let us ask ourselves: what new insights did we get from string theory in
recent years that could shed new light on this question? Clearly the most
striking progress that has been made is the growing evidence of an intimate
connection between quantum phenomena in gauge theory and classical aspects
of gravity. Famous recent examples are D-branes, black hole entropy
counting, matrix theory, and the AdS/CFT correspondence. At a deeper level
these fundamental breakthroughs in string theory -- and even less recent
ones such as the Green-Schwarz anomaly cancellation -- all seem to go back
to the old observation that open string loops are dual to closed string
propagators. It is a logical possibility that a quantum gauge theory/
classical gravity connection of this kind also exists in our low energy world, 
and plays a role in a cancellation mechanism for the cosmological constant.

\bigskip

\newcommand{\SSS}{{\cal S}_0}
\section{The Holographic Correspondence and the Hamilton-Jacobi equation.}

The clearest statement about the duality between gauge theory and gravity is
made in the framework of the AdS/CFT correspondence \cite{ads}.
The statement is: in the strong coupling and large $N$ limit certain
4-d gauge theories have a dual descriptions in terms of 5-d supergravity
theory defined on a space with a 4-d boundary. One of the main
ingredients in this correspondence is the identification of the
classical supergravity action, with specified boundary values for the
metric $g_{\mu\nu}(x)$ and the scalars $\phi^i(x)$, with the generating
function $\SSS$ of gauge theory correlators of gauge invariant
observables ${\cal O}_i$.
\begin{equation}
\addtolength{\abovedisplayskip}{\extraspaces} \addtolength{%
\belowdisplayskip}{\extraspaces} \addtolength{\abovedisplayshortskip}{%
\extraspace} \addtolength{\belowdisplayshortskip}{\extraspace} \Bigl\langle \,
{\cal O}_{i_1}(x_1)\ldots {\cal O}_{i_n}(x_n) \,\Bigr\rangle = {\frac{1}{{\sqrt
{g}}}}\, {\frac{\delta \ \ }{\delta\phi^{i_1}(x_1)}}\ldots{\frac{1}{{\sqrt
{g}}}}\, {\frac{\delta\ \ }{\delta\phi^{i_n}(x_n)}}\, \SSS\, [\,
\phi\, ,\, g\, ].
\end{equation}
It is a standard fact, well known in classical mechanics, that the
classical action as a function of the boundary values satisfies the
Hamilton-Jacobi equation. In a theory with gravity this is a bit more
subtle, since in that case the equations of motion imply that the
Hamiltonian vanishes. So instead of a H-J equation there is a H-J constraint.
Generically, the H-J equation is obtained by replacing the canonical
momenta by the derivatives of the action with respect to the conjugate
variables. To write the H-J constraint we introduce the bracket notation
\begin{equation}
\addtolength{\abovedisplayskip}{\extraspaces} \addtolength{%
\belowdisplayskip}{\extraspaces} \addtolength{\abovedisplayshortskip}{%
\extraspace} \addtolength{\belowdisplayshortskip}{\extraspace} \,
\Bigl\{\SSS,\SSS\Bigr\}\equiv \frac {1}{\sqrt{g}}\left({{\frac{1}{3}}}
\Bigl(g^{\mu\nu}\frac{\delta \SSS }{\delta g^{\mu\nu}}\Bigr)^2 -\frac{\delta \SSS}{\delta
g^{\mu\nu}} \frac{\delta \SSS}{\delta g_{\mu\nu}} -\frac {1}{2} {%
\frac{\delta \SSS}{\delta \phi^i}} {\frac{\delta \SSS}{\delta
\phi_i}}\right)
\label{bracket}
\end{equation}
The variations are all with respect to the fields at the same point $x$.
Hence, the bracket -- which can also be defined for two different actions ${\cal S}_1$ and ${\cal S}_2$ --
is a local density on space-time. In this notation the H-J constraint reads
\begin{equation}
\addtolength{\abovedisplayskip}{\extraspaces} \addtolength{%
\belowdisplayskip}{\extraspaces} \addtolength{\abovedisplayshortskip}{%
\extraspace} \addtolength{\belowdisplayshortskip}{\extraspace} \,
\Bigl\{\SSS,\SSS\Bigr\} =\sqrt{g}\Bigl[V(\phi) +R + \frac{1%
}{2} \, (\nabla\phi)^2\Bigr]
\label{ww}
\end{equation}
The right hand side represents the local lagrangian density of the 5-d gravity
theory with potential $V$ truncated to 4-d fields.
The H-J equation (\ref{ww}) determines the behavior of $\SSS$ under
rescalings of the metric, and thus may be interpreted as an RG-flow
equation. Indeed, it reduces to the standard Callan-Symanzik equation in 
a certain limit \cite{one}, and it is in full accordance with earlier 
ideas on the holographic RG-flow \cite{rg}.
The expression (\ref{ww}) has obvious similarities with the 4d Einstein action.
It has, however, a different interpretation. The potential $V(\phi)$ 
represents the vacuum energy contribution due to
an infinitesimal RG-step. The other terms are interpreted in a similar way as 
modifications of Newtons constant and the scalar coupling constants.

The H-J equation is derived from the classical 5-d Einstein
equations, and hence it is limited to certain special 4-d gauge 
theories for which this approximation is reliable.
Now let us for the purpose of this note
consider a Universe described by such a gauge theory, but then coupled
to gravity. Is there something that the H-J constraint could teach us about
the cosmological constant in that Universe? The action $\SSS$ is the 
effective action induced by integrating out the gauge theory, and has to be 
added to the usual Einstein action possibly with cosmological term.
The problem of the cosmological constant arises because there is no
immediate reason why the cosmological term contained in the
Einstein action would cancel the vacuum energy of the gauge theory.
The two actions usually don't know much about each other. In this note
we want to argue that there actually is a natural relation between
the Einstein action and the effective action of the gauge theory.
To motivate the proposed relation I will use an analogy with
the classical relativistic point particle. So let us, as an inter-mezzo,
briefly discuss some facts about the H-J equation for point particles.

\bigskip

\section{Inter-mezzo: Hamilton-Jacobi theory of the point particle.}

\noindent
The non-relativistic Hamilton-Jacobi equation may be regarded as the
WKB approximation to the Schr\"odinger equation, in which one writes the
wave-function as $\psi=e^{{\frac{i}{\hbar}} S}$ and keeps only the leading
order in $\hbar$. For a relativistic point particle one can do the same for
the Klein-Gordon equation. Hence, 
the H-J equation for a free relativistic
point particle takes the form
\be
\label{HJP}
{\frac{1}{c^2}}
(\partial_t S_0)^2-(\partial_x S_0)^2 =m^2c^2
\ee
Once the classical action $S_0(t,x)$ is known as a function of $x$ at a 
given time $t$, one can use the H-J equation to determine it for later
or earlier times. The solution is not unique, but depends on a choice of
boundary conditions. For example, the classical action for a trajectory from 
$(x,t)$ to $(y,T)$ is
\be
S_0(t,x)=-mc \sqrt{c^2(T-t)^2-(y-x)^2}.
\ee
One easily verifies that it satisfies the H-J equation.
It contains a rest-energy part that diverges in the 
non-relativistic limit $c\to\infty$; we have $S_0\sim -mc^2
(T-t) + m{(y-x)^2/2(T-t)}$. We will find that this rest-energy part plays an 
analogous role as the vacuum energy of the 4d gauge theory. 

The coordinates $x$ and $t$ represent the starting point of a trajectory, and 
should not be regarded as variables to which the action has to be extremized:
indeed, putting the variations of the action $S_0$ to zero would lead to 
nonsensical equations of motion. To obtain an action with a sensible equation 
of motion, we should add the action of another trajectory which 
has $(x,t)$ as its end-point. Let us choose as the additional term the
classical action $S_E$ of a trajectory with a given velocity $v$ and energy 
$E$
\be
S_E(t,x)=-E\left(t- {\frac{v\cdot x}{c^2}}\right).
\ee
To obtain consistent equations of motion, we should demand that 
this action $S_E$ satisfies the {\it same} H-J equation. One easily
checks that this gives the usual relativistic formula for the energy $E$. 

Let us now consider the equations of motion that follow from total action
\be
S(t,x)=S_E(t,x)+S_0(t,x)
\ee
The equations for $x$ are solved by the classical trajectory $x(t)=y+v(t-T)$. 
Somewhat unexpectedly, the equation for $t$ is automatically satisfied once 
we solve the equations for $x$. This can be understood from the fact 
that the total action $S$ satisfies a flow equation. When we insert 
$S_0=S-S_E$ again in the H-J equation, and use the explicit form of $S_E$ we
obtain
\begin{equation}
\addtolength{\abovedisplayskip}{\extraspaces} \addtolength{%
\belowdisplayskip}{\extraspaces} \addtolength{\abovedisplayshortskip}{%
\extraspace} \addtolength{\belowdisplayshortskip}{\extraspace} \gamma(\partial_t+v
\cdot\partial_x)S=-{\frac{1}{2m}}(\partial_x S)^2+ {\frac{1}{2mc^2}}%
(\partial_t S)^2
\end{equation}
This flow equation ensures that when $(x,t)$ is a point on a classical
trajectory, that $(x+v\epsilon,t+\epsilon)$ also is a point on that same
classical trajectory.

Finally, from the explicit expressions for $S_0$ and $S_E$ it follows
that when we take the limit $c\to \infty$ while keeping $v$ fixed, the 
term $mc^2t$ in $S_0$ is canceled by the leading rest-energy part of $E$. 
This cancellation is automatically guaranteed by the fact that 
both actions satisfy the same H-J equation. It is this kind of 
cancellation that we also are 
\bigskip

\section{The Einstein action and the RG flow equations.}

\noindent
I will now repeat the same steps for the H-J equation of 5d-gravity.
First let us identify the corresponding features of equation (\ref{HJP})
and the more elaborate H-J equation (\ref{ww}) of the 5d gravity theory.
The analogue of the time $t$ is obviously the conformal mode of the 
metric, while the coordinates $x$ represent the couplings $\phi$ and
the off-diagonal modes of the metric. The role of the speed of light
$c$ is played by the (square root of) the potential $V$. 
This potential represents
the 5-d cosmological constant, and is also related to the
4d vacuum energy contained in $\SSS$.

The magnitude of the vacuum energy term in $\SSS$, which is the 
analogue of the energy term for the point particle, is restricted by the 
H-J equation. It is expected to be of the order
$m_s^4\sqrt{V}$, where $m_s$ denotes the cut-off scale\footnote{The cut-off scale does not appear explicitly
in our formulas, because it has been absorbed in the metric through a substitution $g_{\mu\nu}\to m_s^{-2} g_{\mu\nu}$.} 
The (square of) $m_s$, which we identify with the
string scale, plays the role of the mass $m$ in the point-particle analogy.
So removing the cut-off is analogous to taking the infinite mass limit.

The large vacuum energy contained in the action $\cal S_0$ is not yet a 
point of concern, since the action $\SSS$ is not an action that has to 
be extremized, in particular not with respect to the conformal mode of 
the metric. To obtain sensible field equations we do like in the point 
particle case, and add an action ${\cal S}_E$ of the form
\begin{equation}
\addtolength{\abovedisplayskip}{\extraspaces} \addtolength{%
\belowdisplayskip}{\extraspaces} \addtolength{\abovedisplayshortskip}{%
\extraspace} \addtolength{\belowdisplayshortskip}{\extraspace} {\cal S}_E=
\int \sqrt{g}\Bigl(U(\phi)+ {1\over \kappa^2(\phi)}
\left[R+ (\nabla\phi)^2\right]+\ldots\Bigr)
\end{equation}
We denote this action by ${\cal S}_E$, since we will indeed identify it with
the Einstein action. The geometric interpretation of ${\cal S}_E$ in the 
AdS/CFT context will be explained below. For the moment we just follow 
what we did in the point-particle case, and require that ${\cal S}_E$ 
satisfies the same H-J equation as $\SSS$. Thus we have
\begin{equation}
\addtolength{\abovedisplayskip}{\extraspaces} \addtolength{%
\belowdisplayskip}{\extraspaces} \addtolength{\abovedisplayshortskip}{%
\extraspace} \addtolength{\belowdisplayshortskip}{\extraspace} \Bigl\{{\cal S%
}_E,{\cal S}_E\Bigr\}= \sqrt{g}\Bigl[V(\phi) +R + \frac{1%
}{2} \, (\nabla\phi)^2\Bigr].
\label{HJC}
\end{equation}
where we again used the bracket notation (\ref{bracket}).
This equation produces certain relations between the 4-d potential $U$, Newton's constant $\kappa$, and
the 5-d potential $V$. Following the same argumentation as before 
we construct the total action
\be
{\cal S}=\SSS+{\cal S}_E
\ee
The fact that $\SSS$ and ${\cal S}_E$ both satisfy the same H-J equation
is now expected to lead to cancellations between the vacuum energy terms,
just like the rest-energies canceled for the point-particle.  
We will return to this point below.

The action ${\cal S}$ also satisfies a flow equation. By inserting
$\SSS={\cal S}-{\cal S}_E$ in the H-J equation we get
\be
 2\Bigl\{{\cal S}_E,{\cal S}\Bigr\}=\Bigl\{{\cal S},{\cal S}\Bigr\}
\ee
where we used the bracket notation for the case of 
two different actions. Just like in the point particle case we want to 
identify the left-hand-side with the flow of the action ${\cal S}$.
To make the correspondence with an RG-flow equation more explicit we
introduce the quantities
\begin{equation}
\addtolength{\abovedisplayskip}{\extraspaces} \addtolength{%
\belowdisplayskip}{\extraspaces} \addtolength{\abovedisplayshortskip}{%
\extraspace} \addtolength{\belowdisplayshortskip}{\extraspace}
{\frac{1}{\sqrt{g}}} \frac{\delta{\cal S}
_E}{\delta g^{\mu\nu}}=\gamma(3g_{\mu\nu}-{1\over2}\beta_{\mu\nu})
\end{equation}
with $\beta_{\mu\nu}$ traceless and
\begin{equation}
\addtolength{\abovedisplayskip}{\extraspaces} \addtolength{%
\belowdisplayskip}{\extraspaces} \addtolength{\abovedisplayshortskip}{%
\extraspace} \addtolength{\belowdisplayshortskip}{\extraspace} %
\frac{1}{\sqrt{g}}{\delta{\cal S}_E\over \delta \phi^i}=\gamma\beta_i
\end{equation}
The quantities $\gamma$, $\beta_{\mu\nu}$ and $\beta_i$ represent the flow
velocities of the metric and the scalar fields under the RG-transformations.
Note that the off-diagonal mode of the metric also has been given
a beta-function $\beta_{\mu\nu}$. This means that the geometry in general may
depend on the RG-scale in a way that differs from just a rescaling of 
the metric.
In terms of these quantities the flow equation becomes
\begin{equation}
\addtolength{\abovedisplayskip}{\extraspaces} \addtolength{%
\belowdisplayskip}{\extraspaces} \addtolength{\abovedisplayshortskip}{%
\extraspace} \addtolength{\belowdisplayshortskip}{\extraspace} \gamma
\Bigl(
2g^{\mu\nu}{\frac{\delta\ }{\delta g^{\mu\nu}}}- \beta^{\mu\nu}{\delta\ \over
\delta g^{\mu\nu}}-\beta^i{\frac{\delta\ }{%
\delta\phi^i}}\Bigr){\cal S} =\Bigl\{{\cal S},{\cal S}\Bigr\}
\end{equation}
here the right hand side represent a kind of anomaly term, which however
vanishes on shell. This equation is analogous to the point particle flow 
equation and guarantees that the solutions to the field equations describe 
RG-trajectories rather then particular field configurations. 
Namely, if $(g_{\mu\nu},\phi^i)$ solves the field equations then 
$(g_{\mu\nu}+\delta_\epsilon g_{\mu\nu},\phi^i+\delta_\epsilon\phi^i)$ 
also represents a solutions with
\be
\label{flow}
\delta_\epsilon g^{\mu\nu}=\epsilon\left(-2 g^{\mu\nu}+\beta^{\mu\nu}\right)
\qquad\qquad \delta_\epsilon \phi^i=\epsilon \beta^i(\phi)
\ee
where $\epsilon$ is an arbitrary infinitesimal space-time dependent function.
Note that in this framework it is natural to consider space-time dependent
RG-flow equations, which reflects the fact that the formalism 
descents from 5d gravity.

The bracket of ${\cal S_E}$ with itself can be expressed entirely in the $\gamma$-factor and
the beta-functions $\beta_i$ and $\beta_{\mu\nu}$. We have
\begin{equation}
\addtolength{\abovedisplayskip}{\extraspaces} \addtolength{%
\belowdisplayskip}{\extraspaces} \addtolength{\abovedisplayshortskip}{%
\extraspace} \addtolength{\belowdisplayshortskip}{\extraspace} 
\Bigl\{{\cal S%
}_E,{\cal S}_E\Bigr\}=12\gamma^2
\left( 1 - {\frac{1}{48}} \beta^{\mu\nu} \beta_{\mu\nu} - {\frac{1}{24}}
\beta_i^2\right) 
\end{equation}
Thus the H-J constraint (\ref{HJC}) gives a condition on the quantities on 
the right hand side. 
The similarity with the point-particle case should be obvious.
In fact, the form of this relation suggests that there exist a Lorentz
type of symmetry that acts on the metric, the scalar fields and 
on the beta-functions. The role of this symmetry is not yet clear to us.

\bigskip

\section{The Einstein equations and the cosmological constant.}

The field equations that follow from the total effective action are the
Einstein equation and the scalar field equation
\be
{1\over\kappa^2}\left(R_{\mu\nu}- {
{\frac{1}{2}}}R g_{\mu\nu}\right)+{
{\frac{1}{2}}}
U(\phi)g_{\mu\nu}=
T_{\mu\nu}^\phi +
\langle\, T_{\mu\nu}\, \rangle  \label{einst}
\end{equation}
\begin{equation}
\addtolength{\abovedisplayskip}{\extraspaces} \addtolength{%
\belowdisplayskip}{\extraspaces} \addtolength{\abovedisplayshortskip}{%
\extraspace} \addtolength{\belowdisplayshortskip}{\extraspace}
\nabla\left({2\over\kappa^2} \nabla\phi_i\right)+2{\partial_i\kappa \over \kappa^3}\left[R+(\nabla\phi)^2\right] =
\partial_i U + \langle\, {\cal O}_i\, \rangle  \label{scal}
\ee
where $T^\phi_{\mu\nu}$ denotes the stress tensor of the scalar fields and
$\langle T_{\mu\nu}\rangle$ and $\langle {\cal O}_i\rangle$ are the expectation
values of the stress energy tensor and the operator ${\cal O}_i$
in the gauge theory to which the scalar field $\phi^i$ couples.
As we just mentioned, the flow equation implies that once we have a
solution at a  particular scale there is a solution along the whole
RG-trajectory. Furthermore,
the equations for the conformal mode of the metric are automatically
implied once we have solved the other field equations.

These other equations do not depend on the cosmological constant, and naively
one may think therefore, that one can simply shift the cosmological constant
and still have a solution. This is not a valid procedure, however, since
the solution of the other field equations determines the cosmological
constant,
and not visa versa -- see also \cite{weinberg}. To explain this point in more 
detail, let us have a closer look at the trace of the Einstein equation
\be
{1\over\kappa^2 } R =
2U(\phi)-\langle\, T^\mu{}_\mu\, \rangle
\end{equation}
where we have taken the fields $\phi_i$ to be space-time independent.
Here the two terms on right-hand-side together represents the cosmological
constant. At first sight there appears to be no reason to suspect a
 relation  between these two terms. However, in our approach there is
additional information in the form of the H-J equation. It is important
to realize that the H-J equation is not an equation of motion, but merely
a condition on the variations of both the $\SSS$ and ${\cal S}_E$ part
of the action. So we may consider this condition for any field configuration,
including a preferred one, such as flat space with constant scalar fields.
In this case we find the following identities for $U$ and
$\langle T^\mu{}_\mu\rangle$ from the H-J equations
\be
2U(\phi) = \sqrt{12V(\phi)+6(\partial_i U)^2}
\end{equation}
and
\be
\langle T^\mu{}_\mu \rangle=\sqrt{12V(\phi)+6\langle {\cal O}_i\rangle^2}\qquad\quad\mbox{for $\ R=0\ $
and $\ \nabla\phi^i=0$.}
\end{equation}
To derive this result we did not make use of the equations of motion.
But now we observe that when we do insert the equation of motion for $\phi$,
which for constant fields reads
\be
\partial_iU +\langle {\cal O}_i\rangle =0
\ee
that both terms indeed cancel!
However, this is no surprise, since we
already knew that once the other field equations are solved that also the 
trace of the Einstein equation holds\footnote{This assumes 
that a solution exists. There are
arguments that in a theory with an invariance of the type (\ref{flow})
not all equations can be satisfied, unless one fine tunes the
parameters in the action \cite{weinberg}. I believe that these arguments do 
not apply here, since it would mean that there does not exist a solution to
the 5-d Einstein equations for generic boundary conditions.} . 
Indeed, in a similar way one could  argue that there always is a
solution to the field equations for any value of the cosmological constant.
This is natural from the point of view of the local flow equation, because
metrics with different constant curvature are related by a conformal rescaling.
In other words: there is whole family of solutions, all related by the local
RG-flow, among which there is in general one that looks like flat space.
So even though we have not resolved the cosmological constant problem
completely,
we did make important progress: we have now the possibility of having a flat
space solution. This leaves the question of why Nature would choose that
particular solution and not some other value for the cosmological constant.
The answer to this question requires more physical input, such as the
boundary conditions in the UV and in the IR. I will not say much about that 
here, and refer to \cite{two} for more details on this point.

\bigskip

\section{Geometric interpretation in terms of
Warped Compactifications.}

The equations that we described have a natural geometric interpretation as a
compactification of the holographic AdS/CFT correspondence. In the usual
non-compact AdS/CFT set-up  gravity is decoupled from the dual boundary theory,
because the graviton modes that extend all the way to the UV boundary are not
normalizable. However, as pointed out in \cite{rs}, this situation changes
as soon as one truncates the $UV$ region by introducing a brane-like
structure (a ``Planck brane''). As a result there will exist normalizable
modes of the 5-d metric that couple as 4-d gravitons to the dual gauge
theory with the correct gravitational strength \cite{rs,hv}.

Instead of cutting off the UV-region with a real Planck Brane one can
alternatively consider the holographic region as a part of a bigger warped
compactification manifold with geometry
\begin{equation}
\label{warp2}
\addtolength{\abovedisplayskip}{\extraspaces} \addtolength{%
\belowdisplayskip}{\extraspaces} \addtolength{\abovedisplayshortskip}{%
\extraspace} \addtolength{\belowdisplayshortskip}{\extraspace} ds^2 = a^2(r)
\eta_{\mu\nu}dx^\mu dx^\nu \, + h_{mn}(r) dr^m dr^n.  \label{warp}
\end{equation}
where the fifth coordinate has become one of the coordinates of a
compact 6-manifold $K_6$. This kind of warped geometry naturally arises in a
class of type IIB string compactifications based on type IIB with a large
number of $D3$ branes. These D3-branes wrap the 4-d uncompactified world,
and are localized as point-like objects inside the $K_6$. When a relatively
large number of D3-branes are located near or at the same point one typically
gets a warped geometry of the kind (\ref{warp2}), in which warp factor
in the neighbourhood of that point essentially depends only
the radial distance $r$ from the D3-branes. By cutting the warped
compactification manifold at some finite value of $r$ one splits
the manifold in two parts: one part, which we call $\Sigma_0$
describes the near horizon region of the D3 branes, and the
other part $\Sigma_E$ the rest of the compactification manifold.
\figuur{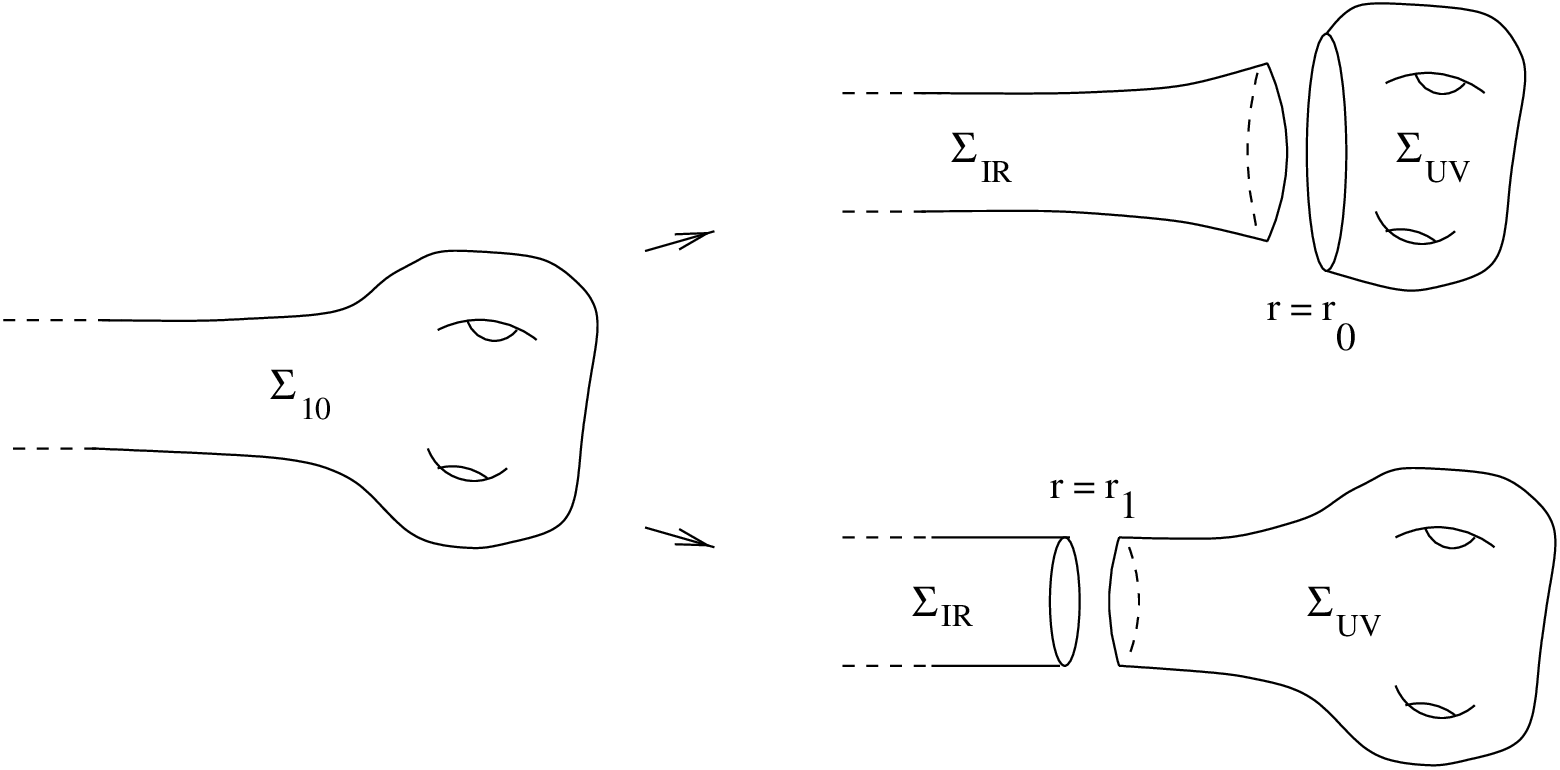}{8cm}{The warped compactification manifold is split up in two
parts $\Sigma_0$ and $\Sigma_E$ by cutting at a radial
location $\ttau$ close to the end of the tube, or at a location $\ttau =%
\ttau_1$ inside the tube.}
In this geometric picture the two actions $S_0$ and $S_E$ can be
identified with the classical supergravity action of that part of the field
configuration that extends in $\Sigma_0$ or in $\Sigma_E$.

Let me now explain how  the results in the previous
subsection fit in this geometric picture. The 4d metric and fields $\phi_i$ 
represent the values of the higher dimensional metric and field $\phi_i$ at a 
particular slice in the geometry. The flow equation is a reflection of the 
arbitrariness in the choice of that slice. The generic existence of a flat 
space-time solution is due to the fact that it is possible to choose slices 
for which the 5d cosmological term is entirely by the expansion 
(or contraction) of the warp factor $a(r)$ and the flow velocities 
of the scalar fields. The effective 4d cosmological constant is
\be
\label{lambda}
\Lambda= a^2V(\phi)-12{\dot{a}^2} +
{1\over 2} a^2 \dot\phi_i^2
\ee
where the dot denote differentiation with respect to the radial coordinate.
A cancellation between these three terms seems to require fine tuning.
But one easily checks that, for space-time independent fields $\phi^i$,
once these terms cancel at one value of $r$, they remain equal at other
values of $r$ by the equations of motion. Mathematically this calculation
is identical -- except for the change from 4 to 5 dimensions and from a 
time to a radial variable-- to the cosmological evolution of an 
inflationary Universe at critical density ($k=0$). Indeed, it 
follows trivially from the Friedman equations that a Universe with a 
flat spatial geometry will stay flat, even if some phase-transition occurs.
One can also see this fact in the Hamilton-Jacobi formalism.
The flow velocities of $a$ and $\phi^i$ are given by
$$
\dot{a}={1\over 6}U \,a\qquad\ \mbox{and}\ \quad \dot{\phi}_i=\partial_iU,
$$
while through the H-J equation the potential $V$ is expressed as
$$
V={1\over 3}U^2-{1\over 2}(\partial_iU)^2
$$
Adding the three contributions in (\ref{lambda}) indeed gives a zero total 
result. This is precisely the condition that was noted to be sufficient to \
have stable brane-like structures with vanishing curvatures \cite{townsend}. 
But again, in a similar way one can also construct solutions that have a 
constant non-zero curvature. The equations again look very much like the 
Friedman equations for an inflationary Universe with non-zero spatial 
curvature $k\neq 0$. Except now the curvature is that of the whole Universe 
itself.

\bigskip

\section{Discussion and Conclusions}

The main result that we derived, the flow equation for the effective action, 
is based on the assumption that the 5d gravity equations are applicable and 
describe the radial propagation of the effective action. This requires that 
the gauge theory must have a large coupling constant, and large enough gauge 
group, and furthermore our energy scale should be small enough so that we can 
ignore string and quantum gravity corrections. Outside this low energy, 
strongly coupled, large $N$ regime one should therefore expect to get 
significant modifications of the flow equations. Before discussing what 
we have learned about the cosmological constant, let me describe what I 
expect some of these corrections will look like.

First of all, $1/N$ corrections would change the flow equation (\ref{flow}) 
by adding a second term which is expressed as a second order variation of
the action $S$. This is analogous to the transition of the H-J equation to
the Schroedinger equation. Indeed, the flow equation should in that case be 
derived from the Wheeler-DeWitt constraint rather than the H-J constraint. 
Although it is hard to give a precise definition of the WDW equation, 
it does at least formally exhibit the same flow invariance.  
Secondly, at some energy scale we expect that the $\alpha^\prime$ 
corrections in the 5d gravity would also become more important. 
One might try to add some higher derivative terms, but propably
a better way to do it is to go to a non-local form of the equation like
a loop equation \cite{polyakov} or use string field theory. 
Finally, when quantum gravity effects in 4d gravity become important 
one has to go to the full string language, and think about the gauge 
theory as open strings and the gravity part as closed strings. 
Also in this case one expects to have a flow equation that relates the 
generating function of open string amplitudes to a part of the closed 
string action \cite{hv2}.

Finally let us return to the issue of the cosmological constant.
The ideas presented in this note shed new light on the issue of the 
cosmological constant, but clearly more work needs to be done to provide 
a solution to the problem. In geometric terms I have shown that by embedding 
our 4d world as a slice in a higher dimensional space that there is a natural 
way in which the vacuum energy that is generated by phase transition is 
canceled by an ever decreasing (towards the IR) warp factor. Via the H-J 
formalism this translates to the statement that the potential $U$ is 
compensated by the trace of the stress energy tensor at all scales, 
once this is the case at one particular scale. Both these statements 
only hold if one restricts to a particular conformal frame of the metric, 
or in terms of the warped geometry, to a particular slicing.

Thus, in the presented scenario the remaining open question is to explain 
why these particular slicings or conformal frames are the right ones. In a 
specific warped compactification it is possible that one can identify such 
a preferred slicing due to the presence of domain wall like structures that 
represent a physical transition at a certain given scale, like a Planck Brane.
It is also possible that a better understanding of 5d holography -- 
and possibly 4d holography -- will reveal an underlying principle that 
can be used to answer this questions. I will leave this question for 
future research.

\bigskip

\section{Acknowledments} I'd like to thank the organizers for
a stimulating, well-organized  and memorable Strings '99 conference.

\end{document}